\documentstyle[preprint,aps,pre,epsf]{revtex}

\begin{document}
\draft
\preprint{draft}
\title{ Random field Ising systems on a general hierarchical lattice:
        Rigorous inequalities}
\author{Avishay Efrat and Moshe Schwartz}
\address{School of Physics and Astronomy, Tel Aviv University, Ramat Aviv,
         Tel Aviv 69978, Israel}
\date{\today}
\maketitle
\begin{abstract}
Random Ising systems on a general hierarchical lattice with both,
random fields and random bonds, are considered. Rigorous inequalities
between eigenvalues of the Jacobian renormalization matrix at the pure
fixed point are obtained. These inequalities lead to upper bounds on the
crossover exponents $\{\phi_i\}$.
\end{abstract}
Despite the many years of research and the large number of researchers
working on the subject, the study of critical behavior of random
systems has led to only few exact results. On the other hand, some of
these results \cite{ss85,s91} played an important role, especially in
the context of the random field problem. In a recent study \cite{es00}
we considered a random bond Ising system on a general hierarchical
lattice, where the renormalization group (RG) transformation is exact
\cite{kg81}, and obtained inequalities concerning the eigenvalues
$\{\lambda_i\}$ of the Jacobian renormalization matrix, at the pure
fixed point. The purpose of the present study is to show that similar
inequalities can be obtained if random fields are included. In
contrast to the case of random bonds and zero fields, correlations are
now generated by the renormalization flow. Nevertheless, it appears
that these correlations are, first, confined to the fields so that the
distribution of bonds is left uncorrelated, and second, restricted to
nearest-neighbor (nn) correlations. It is important to emphasize that
these short-ranged field correlations are generated by the RG
transformation even if one assumes no correlations to begin with, and
that the range of correlations does not increase under the
transformation. Our results are relevant to real lattices,
since some approximate RG schemes on real lattices are
in fact exact RG schemes on hierarchical lattices (Migdal-Kadanoff
\cite{m75,k77} and others \cite{tmss90}) and since it is believed that
the critical behavior of an Ising system on a real lattice can be
mimicked by that behavior on a properly chosen hierarchical lattice
\cite{bo79,bz79,kk84,t85}.

\begin{figure}
\centerline{
\epsfysize 4.08cm
\epsfbox{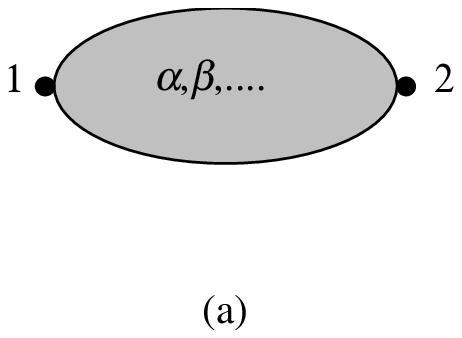}
\qquad
\epsfysize 6cm
\epsfbox{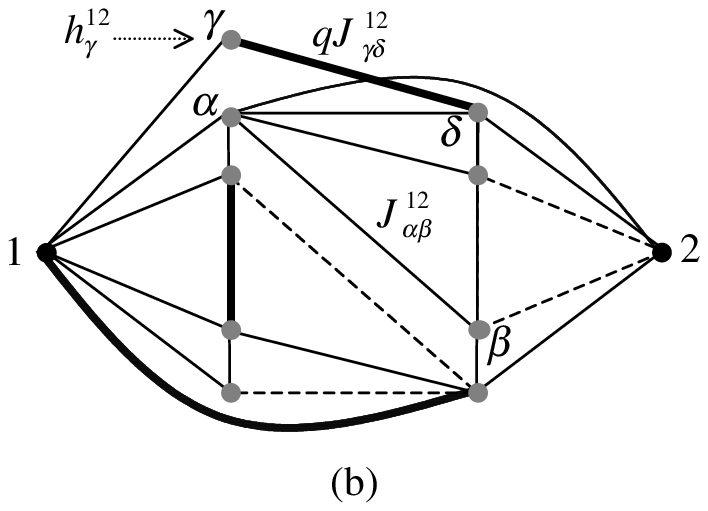}}
\vspace{5mm}
\caption{A general hierarchical lattice is described schematically. In
  (a), the shaded area consists of a set of lattice points,
  $\alpha,\beta,\dots$, where some of the pairs are joined. In (b), a
  typical shaded area is represented. The solid lines are bonds to be iterated in
constructing the lattice, while the dashed ones are not to be
iterated. The bold lines represent the possibility for some bonds to
be strengthened, multiplied by some constant.}
\label{GeneralHLFig}
\end{figure}

We consider a general hierarchical lattice described schematically in
Fig. \ref{GeneralHLFig}. The shaded area shown in (a) consists of a
set of lattice points where some of the pairs are joined. In (b), a typical shaded
area is represented. The solid lines are bonds to be iterated in
constructing the lattice while the dashed ones are not to be
iterated. The bold lines represent the possibility for some bonds to
be strengthened, namely, multiplied by some constant. The random Ising
system is represented by the dimensionless Hamiltonian
\begin{equation}
-{\cal H}=\sum_{( i,j)}J_{ij}\sigma_i\sigma_j+\sum_i
h_i\sigma_i,
\label{SiteFieldHam}
\end{equation}
where $(i,j)$ refers to connected pairs only. All three types of bonds
of Fig. \ref{GeneralHLFig}(a) then carry a coupling
$J^{12}_{\alpha\beta}$ (for the bonds joining sites $\alpha$ and
$\beta$), while each site carries a field $h^{12}_\alpha$. (Note that
one of the members of the pair $\alpha\beta$ may be either 1 or 2.)

The renormalized couplings and renormalized fields are given by
\begin{mathletters}
\label{RenCouplings}
\begin{equation}
\tilde{J}_{12}=f_J\{J^{12}_{\alpha\beta},h^{12}_\alpha\}
\label{RenCoupling}
\end{equation}
and
\begin{equation}
\tilde{h}_{i}=h_i+\sum_{j=1}^{\tilde{z}_i}
f_h\{J^{12}_{\alpha\beta},h^{12}_\alpha\} \qquad i=1,2 \qquad
\label{RenField}
\end{equation}
\end{mathletters}
respectively, while $\tilde{z}_i$ is the coordination number of the
site $i$ on the renormalized lattice. Both, $f_J$ and $f_h$, depend
only on couplings and fields within the rescaling volume associated
with the pair of sites $(1,2)$ (the shaded area in
Fig. \ref{GeneralHLFig}(a)). Eq. (\ref{RenCoupling}) implies that
$\tilde{J}_{ij}$ and $\tilde{J}_{lm}$ are not correlated if the pairs
$(i,j)$ and $(l,m)$ are not identical. This does not hold for the
renormalized fields. Due to the sum in Eq. (\ref{RenField}) over nn
sites on the renormalized lattice, it is clear that even if there are
no correlations to begin with, correlations are generated by the RG
transformation, between fields on nn sites and fields and couplings on
a site and a bond attached to it. (For example, in
Fig. \ref{GeneralHLFig}(b), the following pairs are correlated:
$(\tilde{h}_{1},\tilde{h}_{2})$, $(\tilde{J}_{12},\tilde{h}_{1})$ and
$(\tilde{J}_{12},\tilde{h}_{2})$.) It is easier to deal with such
correlations by considering a bond-field Hamiltonian of the form
\begin{equation}
-{\cal H}=\sum_{(i,j)}[J_{ij}\sigma_i\sigma_j+h_{ij}(\sigma_i+\sigma_j)],
\label{BondFieldHam}
\end{equation}
in which the random variables are the couplings $J_{ij}$ and the bond-fields
$h_{ij}$. The set of RG transformation equations is now given by
\begin{mathletters}
\label{NRenCouplings}
\begin{equation}
\tilde{J}_{12}=f_J\{J^{12}_{\alpha\beta},h^{12}_{\alpha\beta}\}
\label{NRenCoupling}
\end{equation}
and
\begin{equation}
\tilde{h}_{12}=f_h\{J^{12}_{\alpha\beta},h^{12}_{\alpha\beta}\}.
\label{NRenField}
\end{equation}
\end{mathletters}
Eqs. (\ref{NRenCouplings}), therefore
imply that non of the two couplings $\tilde{J}_{ij}$ or
$\tilde{h}_{ij}$, is correlated with any of the two couplings
$\tilde{J}_{lm}$ or $\tilde{h}_{lm}$, if the pairs $(i,j)$ and $(l,m)$
are not identical.

In terms of the bond-fields, the site-fields are given by
\begin{equation}
h_i={1 \over 2}\sum_{j(i)}h_{ij}.
\label{BondF2SiteFMap}
\end{equation}
where $j(i)$ indicates that the sum is over all sites $j$ connected to
$i$. A similar bond-field Hamiltonian (\ref{BondFieldHam}) was
already used in the past \cite{fbm95,yb97,mb98}, only with the
additional term $\sum_{(i,j)} h^{\dag}_{ij}(\sigma_i-\sigma_j)$. Note
that it is necessary to include the dagger fields only if one assumes that the
site-fields are initially uncorrelated. Since, however, the initial
state of non correlated site-fields is not preserved by RG
transformation and nn correlations between site-fields are developed,
there is no reason to start with uncorrelated fields on the sites.

\begin{figure}
\centerline{
\epsfysize 8cm
\epsfbox{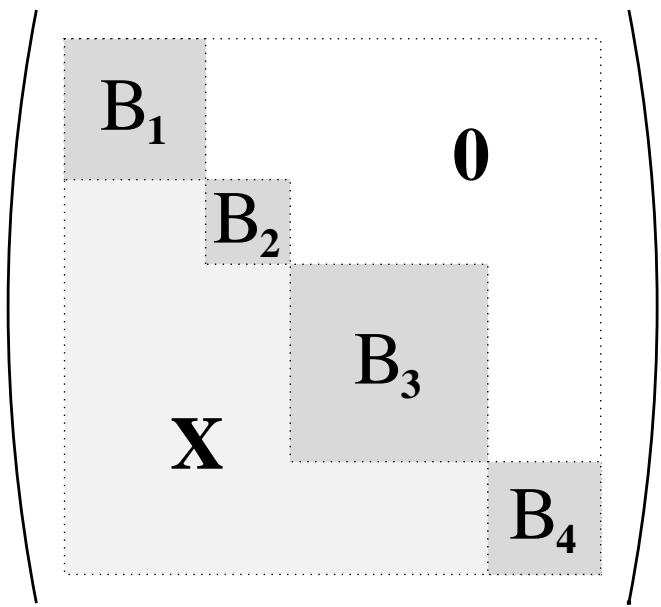}}
\vspace{5mm}
\caption{A general schematic description of a ``block-triangular''
  matrix is shown. ${\bf B}_1,\dots,{\bf B}_4$ represent the blocks
  along the main diagonal. The gray area marked with ${\bf X}$
  indicates the presence of non zero matrix elements while in the
  area marked with $0$, all elements are zero.}
\label{BlockTriangularFig}
\end{figure}

We assume the existence of a ferromagnetic fixed point at
$\{J_{\alpha\beta}\}=J^*$ and $\{h_{\alpha\beta}\}=0$. We denote the
departure of $J_{\alpha\beta}$ from $J^*$ by $\delta J_{\alpha\beta}$
and define the moments as
\begin{equation}
\Gamma_{ij}=\langle (\delta
J_{\alpha\beta})^i(h_{\alpha\beta})^j \rangle.
\label{Moments}
\end{equation}
Clearly at the fixed point $\Gamma_{ij}^*=0$. We will be interested in
the recursion relations of the moments near the pure fixed point,
\begin{equation}
{\tilde \Gamma}_{ij}=G_{ij}\{\Gamma_{lm}\}.
\label{RenMoments}
\end{equation}
The recursion relations above are obtained from the recursion
relations for the local couplings
\begin{mathletters}
\label{ReducExpan}
\begin{equation}
\delta \tilde J_{12}=\sum_{(\alpha,\beta)}\left({\partial f_J
  \over \partial J_{\alpha\beta}}\right)^* \delta J_{\alpha\beta}+
{1\over2} \sum_{(\alpha,\beta)}\left({\partial^2 f_J
  \over \partial J_{\alpha\beta}^2}\right)^*(\delta J_{\alpha\beta}^2)+
{1\over2} \sum_{(\alpha,\beta)}\left({\partial^2 f_J
  \over \partial h_{\alpha\beta}^2}\right)^* h_{\alpha\beta}^2+\dots
\label{ReducExpanJ}
\end{equation}
and
\begin{equation}
\tilde h_{12}=\sum_{(\alpha,\beta)}\left({\partial f_h
  \over \partial h_{\alpha\beta}}\right)^* h_{\alpha\beta}+
\sum_{(\alpha,\beta)}\left({\partial^2 f_h \over \partial
  J_{\alpha\beta}\partial h_{\alpha\beta}}\right)^*\delta
J_{\alpha\beta}h_{\alpha\beta}+\dots,
\label{ReducExpanh}
\end{equation}
\end{mathletters}
where $(\dots)^*$ denotes evaluation at the pure fixed point. Note
that although we are interested only in the expansion of ${\tilde
  \Gamma}_{ij}$ to first order in $\Gamma_{lm}$, we still need, in
principle, orders higher than linear in Eqs. (\ref{ReducExpan})
above. Note also the terms missing in the expansions due to the
different parities of $J$ and $h$. The renormalized moments are given
by
\begin{equation}
{\tilde \Gamma}_{ij}=\sum_{(\alpha,\beta)}\left[\left({\partial f_J
  \over \partial J_{\alpha\beta}}\right)^*\right]^i \left[\left({\partial f_h
  \over \partial h_{\alpha\beta}}\right)^*\right]^j  \Gamma_{ij}+
\sum_{(\alpha,\beta)}\sum_{l,m}A_{lm}^{ij}\Gamma_{lm},
\label{RenMomentsExpan}
\end{equation}
where clearly, in the last sum on the right end side of the above,
$l+m>i+j$. Also the parity of $m$ in the sum must equal the parity of $j$.
The $A_{lm}^{ij}$'s with $l+m>i+j$ always involve derivatives higher
than the first of at least one of the $f$'s. We arrange next the
$\Gamma_{ij}$'s using a single index
\begin{equation}
{\cal G}_k=\Gamma_{ij},
\label{SingleIndex}
\end{equation}
with
\begin{equation}
k={(i+j)(i+j+1) \over 2}+j+1.
\label{kIndex}
\end{equation}
This brings Eq. (\ref{RenMomentsExpan}) into the standard matrix
notation
\begin{equation}
{\tilde {\cal G}}_m=A_{mn}{\cal G}_n.
\label{RenSingleIndex}
\end{equation}
It is not difficult to show that if $k_1$ corresponds to $(i,j)$ and
$k_2$ to $(l,m)$ then $l+m>i+j$ implies $k_2>k_1$. This means that the
matrix $A$ is block-triangular
(Fig. \ref{BlockTriangularFig}). Consider next one of the blocks along
the diagonal of $A$. From the expansions (\ref{ReducExpan}) and
(\ref{RenMomentsExpan}) it follows directly that the only contribution
to $(\delta {\tilde J})^i ({\tilde h})^j$ in $(\delta J)^l (h)^m$ such that
$l+m=i+j$, is the one with $l=i$ and $m=j$. The final conclusion thus
is that the matrix $A$ is triangular, so that its eigenvalues are just
its diagonal terms. The eigenvalues of the Jacobian transformation
matrix are thus
\begin{equation}
\lambda_{ij}=\sum_{(\alpha,\beta)}\left[\left({\partial f_J
  \over \partial J_{\alpha\beta}}\right)^*\right]^i \left[\left({\partial f_h
  \over \partial h_{\alpha\beta}}\right)^*\right]^j.
\label{EigenValues}
\end{equation}
This leads now to a number of interesting inequalities:
\begin{itemize}
\begin{mathletters}
\label{Propo}
\item[(a)]All eigenvalues are positive,
\begin{equation}
\lambda_{ij} \geq 0.
\label{Propoa}
\end{equation}
\item[(b)]All eigenvalues are ordered,
\begin{equation}
\lambda_{i+1,j}<\lambda_{ij} \qquad \text{and} \qquad
\lambda_{i,j+1}<\lambda_{ij}.
\label{Propob}
\end{equation}
\item[(c)]All eigenvalues obey a convexity condition,
\begin{equation}
\lambda_{ij}\lambda_{kj} \geq \lambda_{i+k,j} \qquad \text{and}
\qquad \lambda_{ij}\lambda_{ik} \geq \lambda_{i,j+k}. 
\label{Propoc}
\end{equation}
\item[(d)]All eigenvalues obey
\begin{equation}
(\lambda_{ij})^2 \leq \lambda_{i+k,j+l}\lambda_{i-k,j-l},
\label{Propod}
\end{equation}
where $k=-i,\dots,i$ and $l=-j,\dots,j$.
\end{mathletters}
\end{itemize}

Proof:   In a recent paper \cite{es00} we have considered random bond
Ising systems for which the subset $\{\lambda_{i0}\}$ is
considered. There, we have already proven properties (a)-(c) and our proof here
will follow the same line.

Properties (a) and (c) are proven by showing that
\begin{mathletters}
\label{DerivBiggrZero}
\begin{equation}
{\partial f_J \over \partial J_{\alpha\beta}}(J^*,0)\geq 0
\label{DerivBiggrZeroJ}
\end{equation}
and
\begin{equation}
{\partial f_h \over \partial h_{\alpha\beta}}(J^*,0)\geq 0.
\label{DerivBiggrZeroh}
\end{equation}
\end{mathletters}
We have to consider then the specific transformations generated by
\begin{equation}
-{\tilde{\cal H}}=\ln \mbox{tr}' e^{-\cal H},
\label{RenHam}
\end{equation}
where $\mbox{tr}'$ represents trace only over the subset of spins
$\{\sigma_{\alpha}\}$ internal to the rescaling volume, not
including the external spins $\sigma_1$ and $\sigma_2$. The
renormalized couplings and fields given by Eqs. (\ref{NRenCouplings}),
can now be written in the forms
\begin{mathletters}
\label{RenCompoTr}
\begin{equation}
{\tilde J}_{12}=-{1 \over 4}{\mbox{tr}}_{12}\left[\sigma_1
\sigma_2 {\tilde{\cal H}}\right],
\label{RenCompoTrJ}
\end{equation}
\begin{equation}
{\tilde h}_{12}=-{1 \over
  4}{\mbox{tr}}_{12}\left[(\sigma_1+\sigma_2) {\tilde{\cal H}}\right],
\label{RenCompoTrh}
\end{equation}
\end{mathletters}
where ${\mbox{tr}}_{12}$ indicates trace over the two external spins
$\sigma_1$ and $\sigma_2$. The derivatives of ${\tilde J}_{12}$ with
respect to $J_{\alpha\beta}$ and ${\tilde h}_{12}$ with respect to
$h_{\alpha\beta}$ are thus given by
\begin{mathletters}
\label{Derivs}
\begin{equation}
{\partial{\tilde J}_{12} \over \partial J_{\alpha\beta}}(J^*,0)=
{1 \over 4}{\mbox{tr}}_{12}(\sigma_1\sigma_2) \langle \sigma_\alpha
\sigma_\beta \rangle_{12}
\label{DerivJ}
\end{equation}
and
\begin{equation}
{\partial{\tilde h}_{12} \over \partial h_{\alpha\beta}}(J^*,0)=
{1 \over 4}{\mbox{tr}}_{12}(\sigma_1+\sigma_2) \langle
\sigma_\alpha+\sigma_\beta \rangle_{12},
\label{Derivh}
\end{equation}
\end{mathletters}
where ${\langle \dots \rangle}_{12}$ is the average with respect to
$\cal H$ with $\sigma_1$ and $\sigma_2$ held fixed. In calculating the
above derivatives at the pure fixed point, we use the following
symmetry properties of the system,
\begin{mathletters}
\label{SymmProp}
\begin{equation}
{\langle \sigma_\alpha \sigma_\beta \rangle}_{++}^*={\langle
  \sigma_\alpha \sigma_\beta \rangle}_{--}^*,
\label{SymmPropa}
\end{equation}
\begin{equation}
{\langle \sigma_\alpha \sigma_\beta \rangle}_{+-}^*={\langle
  \sigma_\alpha \sigma_\beta \rangle}_{-+}^*,
\label{SymmPropb}
\end{equation}
\begin{equation}
{\langle \sigma_\alpha \rangle}_{++}^*=-{\langle
  \sigma_\alpha \rangle}_{--}^*,
\label{SymmPropc}
\end{equation}
\begin{equation}
{\langle \sigma_\alpha \rangle}_{+-}^*=-{\langle \sigma_\alpha
  \rangle}_{-+}^*,
\label{SymmPropd}
\end{equation}
\end{mathletters}
to obtain
\begin{mathletters}
\label{FixedPointDerivs}
\begin{equation}
{\partial{\tilde J}_{12} \over \partial J_{\alpha\beta}}(J^*,0)=
{1 \over 2}[{\langle \sigma_\alpha \sigma_\beta \rangle}_{++}^*-{\langle
  \sigma_\alpha \sigma_\beta \rangle}_{+-}^*]
\label{FixedPointDerivJ}
\end{equation}
and
\begin{equation}
{\partial{\tilde h}_{12} \over \partial h_{\alpha\beta}}(J^*,0)=
{1 \over 2}[{\langle \sigma_\alpha+\sigma_\beta \rangle}_{++}^*].
\label{FixedPointDerivh}
\end{equation}
\end{mathletters}
Here the sign indices specifically indicate the state of the spins
$\sigma_1$ and $\sigma_2$ and the $*$ indicates that the average is with
respect to the pure fixed point Hamiltonian
\begin{equation}
-{\cal H}^*=J^*\sum_{(i,j)}\sigma_i\sigma_j.
\label{FixedHam}
\end{equation}
Now, according to the GKS inequalities \cite{g67,ks68}, if all the many-spin
couplings $J_A=h_\alpha,J_{\alpha\beta},\dots$ in a general
Ising system are positive, all the many-spin correlations 
$\langle \sigma_A \rangle=\langle \sigma_\alpha \rangle,\langle
\sigma_\alpha \sigma_\beta \rangle,\dots$ must obey $\langle \sigma_A
\rangle \geq 0$. Using Eqs. (\ref{FixedPointDerivs}), the averages
are taken with respect to the pure ferromagnetic Hamiltonian
(\ref{FixedHam}), where the two external spins of each of the
rescaling volumes, which are held fixed, serve effectively as local
fields. When these effective fields are held both positive, the GKS
inequalities hold, so that
\begin{equation}
\langle \sigma_\alpha \rangle_{++}^* \geq 0 \qquad \mbox{and} \qquad
\langle \sigma_\alpha \sigma_\beta \rangle_{++}^* \geq 0,
\label{GKSIneq}
\end{equation}
which is enough already to prove
Ineq. (\ref{DerivBiggrZeroh}). Ineq. (\ref{DerivBiggrZeroJ}) can be
easily shown to hold using, in addition to the GKS inequalities,
other rigorous inequalities, just recently proven \cite{sem00}, also
concerning the many-spin correlations in general Ising systems. It states that if
all the many-spin couplings $J_A$ are positive again, the absolute value of
all the many-spin correlations $\langle \sigma_A \rangle$ does not
increase when the value of any of the couplings is reduced, taking any
value in the interval $[-J_A,J_A]$.  According to this, it is clear
that under reversal of the $+1$ state of any of the two external
spins, the many-spin correlations can not increase. So, we arrive at
the conclusion that
\begin{equation}
\langle \sigma_\alpha \sigma_\beta \rangle_{++}^* \geq \langle
\sigma_\alpha \sigma_\beta \rangle_{+-}^*
\label{SEMIneq}
\end{equation}
and Ineq. (\ref{DerivBiggrZeroJ}) is also proven. This completes the proof
of properties (a) and (c).

We turn now to property (b). here we need to show that
\begin{mathletters}
\label{DerivLessOne}
\begin{equation}
{\partial f_J \over \partial J_{\alpha\beta}}(J^*,0)< 1
\label{DerivLessOneJ}
\end{equation}
and
\begin{equation}
{\partial f_h \over \partial h_{\alpha\beta}}(J^*,0)< 1.
\label{DerivLessOneh}
\end{equation}
\end{mathletters}
at any finite temperature. But, referring to Eqs. (\ref{FixedPointDerivs})
again, it is clear that
\begin{mathletters}
\label{AvsLessOne}
\begin{equation}
{1 \over 2}[\langle \sigma_\alpha \sigma_\beta \rangle_{++}^* - \langle
\sigma_\alpha \sigma_\beta \rangle_{+-}^*] \leq {1 \over 2}[|\langle
\sigma_\alpha \sigma_\beta \rangle_{++}^*| + | \langle
\sigma_\alpha \sigma_\beta \rangle_{+-}^*|] \leq 1
\label{AvsLessOneJ}
\end{equation}
and that
\begin{equation}
{1 \over 2}[\langle \sigma_\alpha \rangle_{++}^* + \langle
\sigma_\beta \rangle_{++}^*] \leq {1 \over 2}[|\langle
\sigma_\alpha \rangle_{++}^*| + |\langle
\sigma_\beta \rangle_{++}^*|] \leq 1,
\label{AvsLessOneh}
\end{equation}
\end{mathletters}
while the equality sign can hold only at zero temperature. This proves
property (b).

We are left now with property (d). Here we use the more general
definition of a scalar product, $({\bf u},{\bf v}) \equiv \sum_i w_i
u^*_i v_i$ where $\forall i$, $w_i \geq 0$ and the corresponding
Schwartz inequality, which reads $\left( \sum_i w_i u^*_i v_i
\right)^2 \leq \sum_i w_i \left| u_i \right|^2 \sum_j w_j \left| v_j
\right|^2$ (here is the only place where the * represents complex
conjugate). We replace, next, the sum over the single index $i$ with
the double index $(\alpha\beta)$ and identify
\begin{mathletters}
\label{IdentSchwIneq}
\begin{equation}
u_{\alpha\beta} \equiv \left[\left({\partial f_J \over \partial
      J_{\alpha\beta}}\right)^* \right]^r \left[\left({\partial f_h \over \partial
      h_{\alpha\beta}}\right)^* \right]^s
\label{IdentSchwInequ}
\end{equation}
with $r=0,\dots,i$ and $s=0,\dots,j$,
\begin{equation}
v_{\alpha\beta} \equiv \left[\left({\partial f_J \over \partial
      J_{\alpha\beta}}\right)^* \right]^p \left[\left({\partial f_h \over \partial
      h_{\alpha\beta}}\right)^* \right]^q
\label{IdentSchwIneqv}
\end{equation}
with $p=0,\dots,i-r$ and $q=0,\dots,j-l$, and
\begin{equation}
w_{\alpha\beta} \equiv \left[\left({\partial f_J \over \partial
      J_{\alpha\beta}}\right)^* \right]^{i-r-p} \left[\left({\partial f_h \over \partial
      h_{\alpha\beta}}\right)^* \right]^{j-s-q},
\label{IdentSchwIneqw}
\end{equation}
\end{mathletters}
to obtain
\begin{equation}
(\lambda_{ij})^2 \leq \lambda_{i+r-p,j+s-q}\lambda_{i-r+p,j-s+q},
\label{SqrLambdaIneq}
\end{equation}
where we have used the fact that all partial derivatives are real and positive.
All is left now is to denote $r-p=k$ with $k=-i,\dots,i$ and,
similarly, $s-q=l$ with $l=-j,\dots,j$, which completes the proof of property (d).

In addition to that, denoting by $m_J<1$ and $m_h<1$, the maximal values
of $\partial{\tilde J}_{ij}/\partial J_{\alpha\beta}$ and
$\partial{\tilde h}_{ij}/\partial h_{\alpha\beta}$ respectively, we obtain
\begin{equation}
\lambda_{ij}\leq\lambda_{kl}m_J^{i-k}m_h^{j-l},\qquad \mbox{with}
\qquad k=0,\dots ,i \qquad \mbox{and}
\qquad l=0,\dots ,j,
\label{FinRelevInt}
\end{equation}
so that we have also proven that the number of relevant interactions at the
pure fixed point is finite. The only case of which the equality sign
hold is the diamond hierarchical lattice (DHL) \cite{kg81,y88}, where
all bonds are equivalent. From Ineq. (\ref{FinRelevInt}) follows an
inequality for the crossover exponents:
\begin{equation}
\phi_{ij}<1+{(i-1)\ln m_J + j\ln m_h \over \ln \lambda_{10}}<1  \qquad
i+j=2,3,\dots,
\label{CrossIneq}
\end{equation}
where $\phi_{ij}=y_{ij}/y_{10}$, $y_{ij}={\ln
  \lambda_{ij}}/{\ln b}$ and $b$ is the rescaling factor. The
condition for criticality of the pure fixed point is
$\max(\lambda_{20},\lambda_{11},\lambda_{02})<1$, while
else, we expect a
random critical point with a different set of critical exponents. It
is interesting to note that it was just recently shown
\cite{e00} that even for the random bond Ising system, the Harris
criterion for pure criticality \cite{h74}, $\alpha_p<0$ ($\alpha_p$
being the specific heat 
exponent) is equivalent to the obvious requirement, $\lambda_{20}<1$,
only in the special case of the DHL. In
the more general case, it was shown that $\alpha_p\leq\phi_{20}$ so that
the Harris criterion is only a necessary condition for pure
criticality to hold and counter examples where $\alpha_p<0$ and
$\phi_{20}>0$ have been presented. The analogous result for the random
system is $\gamma_p \leq \phi_{02}$ ($\gamma_p$ being the susceptibility
exponent of the pure system) but since $\gamma_p$ turns out always to
be positive, the random field is always relevant at the pure critical point.

We wish to conclude by emphasizing that the
inequalities proven here hold not only for exact RG transformations on
HL's but also for all other renormalization schemes (such as MK scheme
\cite{m75,k77}) in which the renormalized couplings or fields are not
correlated or even in cases where it is clear that the correlations
are not important \cite{rsk80}.

\acknowledgments
We wish to thank D. Andelman for drawing our attention to
Refs. \cite{fbm95,yb97,mb98}.


\end{document}